\documentclass[preprint2,twoside]{hwo}

\usepackage{lipsum}

\input{hwo.h}

\begin{document}

\title{\textbf{\LARGE Potential binary supermassive black holes in Markarian 231 as a scientific case for the spectropolarimeter POLLUX}}
\author {\textbf{\large Julie Biedermann,$^{1}$ Frédéric Marin,$^1$ Coralie Neiner,$^2$ Jean-Claude Bouret,$^3$ }}
\affil{$^1$\small\it Université de Strasbourg, CNRS, Observatoire Astronomique de Strasbourg, UMR 7550, F-67000 Strasbourg, France}
\affil{$^2$\small\it  LIRA, Paris Observatory, CNRS, PSL University, Sorbonne University, Université Paris Cité, CY Cergy Paris University, 5 place Jules Janssen, 92190 Meudon, France}
\affil{$^3$\small\it Aix Marseille Univ, CNRS, CNES, LAM, Marseille, France}




\begin{abstract}

It is now widely accepted that a supermassive black hole (SMBH) resides in the core of nearly every galaxy. Within the context of mergers of galaxies, the central SMBHs from each system should form a gravitationally bound binary system. Given the frequency of galaxy mergers throughout cosmic time, the existence of such binary SMBHs is likely to occur within active galactic nuclei (AGNs). However, identifying them observationally remains highly challenging. The majority of reported candidates rely on indirect evidence rather than unambiguous detection. The quasar Markarian 231 (Mrk 231) stands out because of its distinctive ultraviolet (UV) to optical properties, which includes a strong wavelength dependent polarization and an unusual flux decrease at $\sim$ 2500 \AA. In this work, we investigate the potential of spectropolarimetry as a powerful tool for identifying and characterizing SMBH binaries in AGNs. We have developed a simplified analytical model to reproduce the spectral energy distribution (SED) of the quasar Mrk 231 assuming a binary SMBH geometry, and we compute both its total and polarized emission. The main aim is to test whether two separate emitting sources could explain the unique UV to optical signatures of Mrk 231. Such investigations highlight the importance of a future high-sensitivity UV spectropolarimetric instrument. For instance, the proposed Pollux spectropolarimeter for the Habitable Worlds Observatory (HWO) has the potential to significantly contribute to our understanding of AGNs structure and the role of SMBH binaries in quasars. 
\\
\\
\end{abstract}

\vspace{2cm}

\section{Introduction}
  
AGNs are among the most luminous and energetic objects in the Universe. In the unified model of AGNs proposed by \citet{Antonucci_1993}, the activity of AGN is powered by accretion onto a SMBH, (10$^{6}$ M$_{\odot}$ to 10$^{9}$ M$_{\odot}$). The central SMBH is surrounded by an accretion disk, which is responsible for the majority of the optical and UV emission observed in AGNs. A dusty torus surrounds this central engine, and its presence obscures the inner region depending on the observer's line of sight. The broad line region (BLR) located between the accretion disk and the torus is composed of a dense, rapidly moving gas clouds and is responsible for the presence of broad emission lines observed in the spectra of AGNs. On larger scale, along the polar axis, lies the narrow line region (NLR), a lower density structure, that can extend to kiloparsec scale, and is responsible for narrow emission lines. In certain AGNs, collimated relativistic jets are ejected from the nucleus, introducing further complexity to their multi-wavelength signatures. While the overall structure of AGNs is well described by the unified model, growing observational evidence suggests that in certain cases the central engine may consist of a bound pair rather than a single SMBH. Such binary configuration have the potential to significantly modify the geometry of the nuclear regions, thereby leaving distinct imprints on the spectral and polarization properties of AGNs.  

The quasar Markarian 231 provides a remarkable test case in the context of binary SMBH signatures. Is is among one of the most luminous ultra-luminous infrared galaxies (ULIRGs) in the local Universe and exhibits highly unusual spectral and polarization signatures in the UV-optical range \citep{Thompson_1980}. When moving from the optical to the UV, its total flux declines sharply, with a marked break around $\sim$ 2500 \AA, and then gradually increases at shorter wavelengths \citep{Leighly_2014}. This spectral behavior is atypical when compared with standard quasars, whose UV emission usually decreases more smoothly at shorter wavelengths. Moreover,  the polarization of Mrk 231 shows a similarly distinctive behavior at the same wavelength region where the flux break occurs. The degree of polarization is strongly wavelength dependent, reaching values of around 15~\% near the optical, which is a very high measure in non-blazing quasars, and a systematic rotation of the polarization angle coincides with the variation of the polarization degree \citep{Smith_1995, Goodrich_1994, Smith_2004}. 

In order to explain the prominent flux break around $\sim$2500 \AA, \cite{Yan_2015} suggested that Mrk 231 may contain a binary SMBH system. In their scenario, a massive primary SMBH dominates the optical emission, while a less massive secondary, located within the circumbinary disk, contributes predominantly in the far-UV and could explain the slow rise in the flux at wavelengths shorter than 2500\AA. Furthermore, Mrk 231 is a particularly strong candidate for hosting a binary SMBH system, as this quasar is known to be in an advanced stage of galactic merger \citep{Armus_1994}. The present study is an expansion of this initial idea, with the objective of developing a model for the spectral energy distribution expected from a binary SMBH configuration and investigate if the unique polarization and spectroscopic properties observed in Mrk 231 can also be reproduced.

\section{Analysis}
\subsection{Constructing the SED of a binary SMBHs system}

We developed a simplified analytical model to reproduce the SED expected from a binary SMBH system. As a first step, we examined the scenario of a single SMBH using the standard geometrically thin, optically thick accretion disk described by \citet{shakura_Sunyaev} in which the disk radiates locally as a blackbody. The outer edge of the disk was determined by the Toomre stability criterion, which indicates that the disk becomes gravitationally unstable once the system's accretion rate exceeds a threshold that primarily depends on the SMBH mass. With these components, we successfully reproduced the SED of a single SMBH.   

In a second step, we extended the framework to a binary SMBH configuration in accordance with the scenario suggested by \cite{Yan_2015}. In this picture, the secondary lower-mass SMBH opens a gap in the primary disk and accretes matter from the inner edge of the circumbinary environment, forming a new mini-disk that emits mainly in the far-UV range. The geometry of the system was established by calculating the Hill radius to estimate the inner border of the primary disk, the Roche lobe radius to constrain the outer edge of secondary mini-disk and the Innermost Stable Circular Orbit (ISCO) for the inner radius of the mini-disk. Incorporating these elements, the combined emission of both primary and secondary SMBHs was computed, yielding the SED of the binary system, where the contribution of the less massive secondary SMBH is visible in the far-UV, while the emission of the more massive primary SMBH is dominant in the UV to optical.  

To compare the model with observations of Mrk 231, we compiled multi-wavelength flux measurements together with corresponding polarization data obtained from different observations spanning more than 30 years. These data are shown in Figure~\ref{Biedermann_fig1}, where the aperture of each instrument is color-coded. Additional contributions from the dusty torus and the host galaxy, which dominate in the infrared (IR) and optical emission respectively, were used to construct the total SED of the quasar. We used templates from the SED simulations of \citet{Siebenmorgen_2001} for the torus and the SWIRE library from \citet{Polletta_2007} for the galaxy component. The entire SED covering the far-UV to the IR was constructed by summing the four components with the secondary smaller SMBH, the primary SMBH, the host galaxy, and the dusty torus. The combined model of the total SED is shown by the red curve in the top-left panel in Fig.~\ref{Biedermann_fig1}. 

The model successfully reproduces the observed shape of Mrk 231 data, including the pronounced break around $\sim$ 2500 \AA. In our perspective, this drop in the total flux represents the gap between the contribution of the secondary mini-disk dominating in the far-UV and that of the primary bigger SMBH at longer wavelengths. 

\subsection{Modeling the polarimetric signatures}

Polarization, which is defined as the orientation and the degree of ordering of the oscillation of the electric field of light, provides important information about the geometry of AGNs \citep{Marin_2019}. It is described using the Stokes formalism, which introduced a four-component vector, $S=(I, Q, U, V)$, where $I$ represents the total intensity, $Q$ and $U$ are linear polarization, and $V$ corresponds to the circular polarization. Since there are no circular polarization measurement for Mrk 231, we restrict our analysis to linear polarization. The Stokes parameter $Q$ and $U$ are used to derive two key quantities, namely the degree of polarization $P=\frac{\sqrt{Q^{2}+U^{2}}}{I}$ and the polarization angle $\theta = \frac{1}{2}arctan(\frac{U}{Q})$. The degree of polarization is then multiplied by the total flux in order to determine the polarized flux. Thus for each component of our total SED model, the primary and secondary SMBHs, the host galaxy and the torus, we determined best-fit values for the normalized Stokes parameter : $q_{i}=Q_{i}/I_{i}$ and $u_{i}=U_{i}/I_{i}$. These values were obtained by minimizing the $\chi^{2}$ between the model and the data, providing the most likely polarization for each component. The total polarization signal of the system was subsequently calculated by summing the Stokes contributions from all components. 

\begin{figure*}[ht]
    \begin{center}
    \includegraphics[width=1\textwidth]{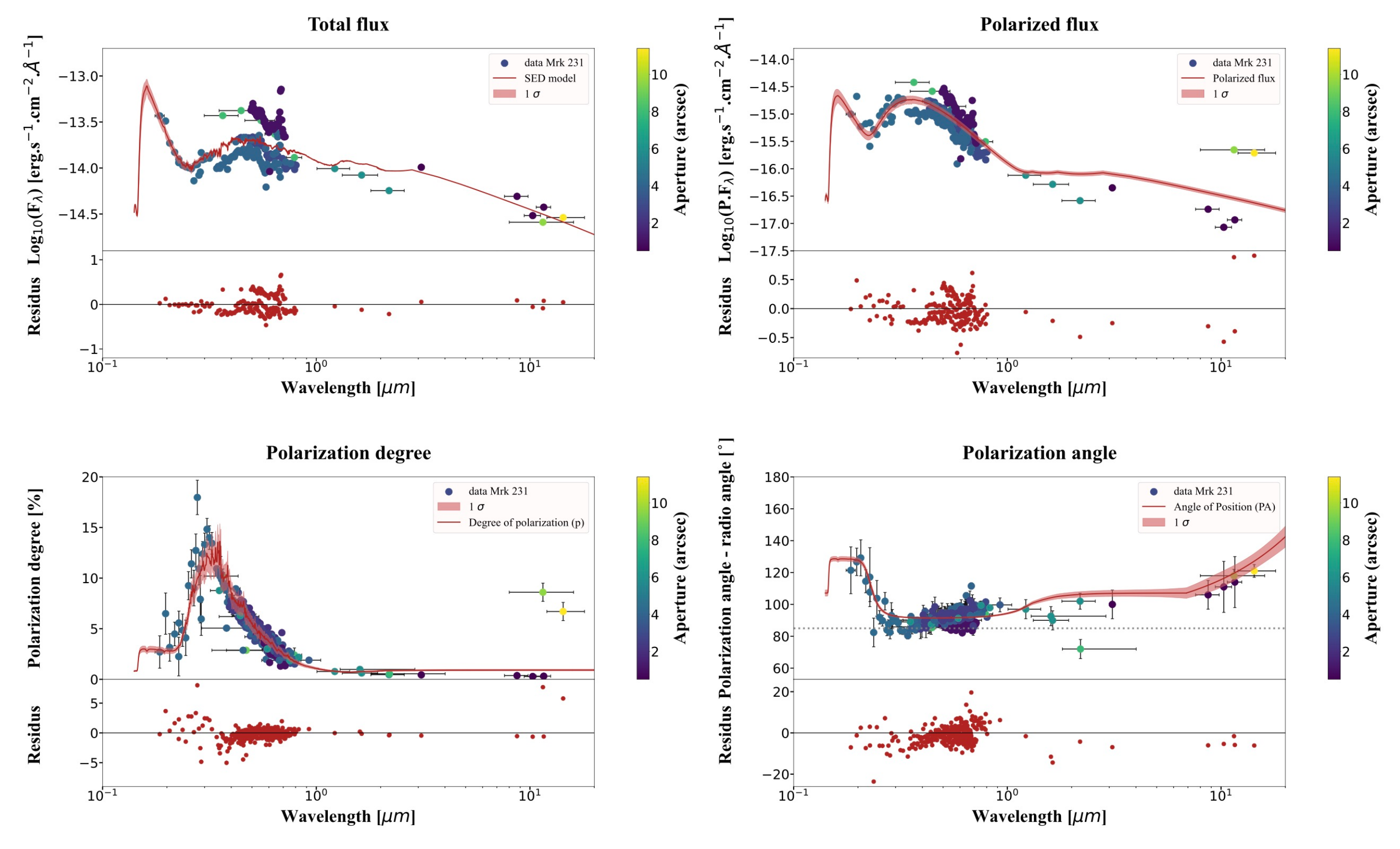}
    \caption{\textit{Top left panel}: Mrk 231 total flux as a function of wavelength. \textit{Top right panel}: Polarized flux, established from the multiplication of the flux and the polarization degree. \textit{Bottom left panel}: Degree of polarization. \textit{Bottom right panel}: Polarization position angle minus the radio position angle \citep{Biedermann_2025}.
    \label{Biedermann_fig1}
    }
    \end{center}
\end{figure*}

The best-fit normalized polarization parameters for each component are presented in Table~\ref{param_pol_final}, and the resulting degree of polarization, polarization angle, and polarized flux are displayed in Fig.~\ref{Biedermann_fig1} (bottom-left, bottom-right, and top-right panels, respectively). The host galaxy contribution was assumed to be unpolarized, consistent with the expectation for stellar emission. The model successfully reproduces the wavelength-dependent variation of polarization observed in Mrk 231. In particular, the optical polarization is influenced by the primary SMBH, allowing the model to produce the observed $\sim$15\%. The model predicts a smooth variation of the polarization angle in the UV, particularly close to the spectral region of the break in the total flux at $\sim$~2500 \AA. The rotation of the polarization angle reflects the transition between the emission of the secondary, lower-mass SMBH in the far-UV, and the optical emission of the primary, more massive one. This thus highlights the role of a binary system in determining the observed polarization properties of Mrk 231. Finally, a secondary bump in the far-UV is visible in the predicted polarized flux, which is directly related to the contribution of the less massive secondary SMBH. In summary, this simple SMBH binary model simultaneously reproduces three independent observables with the total flux, the degree of polarization, and the polarization angle, thus supporting the binary SMBH interpretation as an explanation for the spectropolarimetric signatures of Mrk 231. 

\begin{table}[h]
    \centering
    \caption{Table of the normalized Stokes parameters and best-fit values for the polarization model.}
    \resizebox{\columnwidth}{!}{%
    \begin{tabular}{ c c c }
        \\
        \hline
        Component & q & u\\ 
        \hline
        Primary SMBH & -0.2254 $\pm$ 0.0282 & -0.0475 $\pm$ 0.0009 \\
        Secondary SMBH & -0.0012 $\pm$ 0.0016 & -0.0310 $\pm$ 0.0036 \\
        Torus & -0.0066 $\pm$ 0.0009 & -0.0064 $\pm$ 0.0003 \\
        Host galaxy & 0 & 0 \\
        \hline
    \end{tabular}
    }
    \label{param_pol_final}
\end{table}

However, the model’s ability to confidently confirm the presence of the second bump is severely limited by the weak far-UV coverage of Mrk 231. Therefore, future high-sensitivity spectropolarimetric observations in the 1000–2500 \AA \ range are crucial to test the robustness of these spectropolarimetric signatures and to further assess the plausibility of a binary SMBH scenario in the core of Mrk 231.

\section{Conclusions}

Mrk 231 exhibits very specific UV spectropolarimetric signatures. To interpret the break visible in the total flux, the rotation of the polarization angle and the variation of the polarization degree in the far-UV to UV wavelength range, we developed a simple analytical model based on a binary SMBH geometry, first proposed by \cite{Yan_2015}. Our results provide strong evidence in favor of the binary scenario, as they successfully and simultaneously reproduce three independent observables: the total flux, the polarization degree, and the polarization angle. In addition, this simple model predicts that the less massive secondary SMBH produces a secondary bump in the polarized flux. Such a distinctive polarization characteristic could represent signatures of polarization of a binary SMBH in quasars. However, a confirmation by direct observation is not possible because of the current absence of far-UV spectropolarimetric data. This study highlights how future space-based UV spectropolarimeters could be used to test the presence of binary SMBHs in AGNs and to probe the inner structure of quasar. In particular, significant constraints might be obtained from the next-generation instrument Pollux, a high-resolution UV spectropolarimeter proposed by a European consortium for the Habitable Worlds Observatory (HWO) \citep{Neiner_2025}. With a remarkable spectral resolution (R$\sim$120000 in the far-UV from 100 to 120 $nm$) and covering the 100 $nm$ to 1.8 $\mu m$ wavelength range, Pollux will enable precise measurements of polarized flux exactly where the predicted binary SMBH signatures are the most significant. Such capabilities would pave the way to identifying and characterizing binary SMBHs through their polarization signatures.

{\bf Acknowledgements.} 

Thank you to the organizers of the HWO 2025. 

\bibliography{author.bib}

\end{document}